\documentclass[conference]{IEEEtran}

\IEEEoverridecommandlockouts

\usepackage{amssymb,amsmath,amsthm, amsfonts}
\usepackage{algorithm}
\usepackage{algpseudocode}
\usepackage{array}
\usepackage{graphicx}
\usepackage{cite}
\usepackage{subcaption}
\usepackage[inline]{enumitem}
\usepackage{soul}
\usepackage{caption}
\usepackage[top=0.7in, bottom=1in, left=0.625in, right=0.625in]{geometry}

\usepackage{xcolor}
\DeclareMathOperator*{\argmax}{arg\,max}
\DeclareMathOperator*{\argmin}{arg\,min}
\usepackage{lipsum}
\usepackage{balance}
\usepackage{booktabs}

\allowdisplaybreaks

\newcommand{\ex}{\mathbb{E}}

\newtheoremstyle{italiclabel}
{\topsep}   
{\topsep}   
{\normalfont}  
{}  
{\itshape}  
{.}  
{.5em}  
{}  

\theoremstyle{italiclabel}

\newtheorem{definition}{Definition}

\theoremstyle{normalfont}

\begin{document}
	
	\title{\huge Low-Complexity Tone Injection via Candidate Ranking for PAPR Reduction in OFDM and AFDM Systems}
	
\author{
	\IEEEauthorblockN{Yupeng Zheng$^1$, Ang Li$^2$, Jinfei Wang$^1$, Yi Ma$^1$, and Rahim Tafazolli$^1$}
	\IEEEauthorblockA{$^1$6GIC, Institution of Communication Systems, University of Surrey, Guildford, UK, GU2 7XH\\
		$^1$Email: \{yupeng.zheng, jinfei.wang, y.ma, r.tafazolli\}@surrey.ac.uk}
	\IEEEauthorblockA{$^2$China Telecom Research Institute, Beijing, China\\
		$^2$Email: lia15@chinatelecom.cn}
}
	
	
	\maketitle

\begin{abstract}
Tone injection (TI) is a promising distortionless PAPR reduction technique that incurs no spectral efficiency loss. However, state-of-the-art TI schemes based on random candidate generation or clipping noise spectrum suffer from fundamental limitations in PAPR performance. In this paper, we propose novel TI schemes compatible with both OFDM and AFDM systems. The proposed schemes iteratively update the TI sequence via a candidate ranking procedure guided by time-domain local peaks. This accurately selects effective candidates while achieving a complexity comparable to that of the fast Fourier transform. Depth-first search is further integrated to enhance PAPR performance by exploiting the tree structure of the process. Simulations demonstrate that the proposed schemes achieve over 1~dB PAPR gain over baseline TI schemes at comparable complexity. The gain is consistent across various numbers of subcarriers under controlled per-iteration complexities, confirming a superior performance-complexity trade-off for both OFDM and AFDM.
\end{abstract}

\section{Introduction}
Orthogonal frequency-division multiplexing (OFDM) has been widely adopted in modern wireless standards such as 5G, Wi-Fi, and satellite communications due to its high spectral efficiency and robustness against multipath fading. Recently, affine frequency-division multiplexing (AFDM)~\cite{Bemani2023} has emerged as a promising waveform for high-mobility communications, offering full diversity in doubly-dispersive channels via chirp-based subcarrier modulation. However, both waveforms suffer from a high peak-to-average power ratio (PAPR)~\cite{Rahmatallah2013,Yuan2024}, which drives the power amplifier (PA) into its nonlinear region, causing distortion and degrading system performance.

Various techniques have been proposed to address the PAPR problem, including clipping and filtering, selective mapping, and tone reservation~\cite{Rahmatallah2013}. Among these, tone injection (TI)~\cite{Hwang2001} stands out as a distortionless approach that injects an orthogonal sequence into selected subcarriers, preserving spectral efficiency. However, finding the optimal TI solution requires an exhaustive search over an infinite lattice space, which is computationally prohibitive. Iterative TI schemes have thus been proposed as low-complexity alternatives, broadly falling into two categories. The first~\cite{Chen2010,Jacklin2013,Hou2012} selects the best candidate from a large randomly generated pool, but is computationally inefficient as each candidate requires an FFT to evaluate the time-domain PAPR. The second~\cite{Hou2013,Wang2016} exploits the clipping noise spectrum to restrict TI to the most promising subcarriers. In particular, DS-TI~\cite{Wang2016} iteratively injects signals in order of subcarrier significance, achieving superior PAPR performance at much lower complexity. However, since the clipping noise spectrum varies across iterations, the PAPR reduction of DS-TI is inherently limited.

To overcome the limitations of existing approaches and extend TI to AFDM, this paper proposes a novel TI scheme based on \emph{candidate ranking} (CR-TI). Since adjacent time-domain samples are correlated under oversampling, local peaks serve as proxies for the full signal envelope. Monitoring a small number of local peaks per iteration is therefore far more efficient than tracking the full waveform. Motivated by this, CR-TI ranks all valid candidates based on their weighted contributions to local peak reduction and selects the best candidate to update one subcarrier per iteration. The overall complexity is upper-bounded by $\mathcal{O}(TN^2)$, where $N$ is the number of subcarriers and $T\approx 20$ is the number of iterations. To further reduce complexity, \emph{filtered CR-TI (FCR-TI)} is proposed, achieving an FFT-level complexity by pre-filtering candidates based on the clipping noise spectrum. Both schemes are integrated with \emph{depth-first search (DFS)} to further enhance PAPR performance.

Simulations demonstrate that the proposed schemes achieve $1$~dB PAPR reduction over state-of-the-art TI schemes at comparable complexity. Specifically, CR-TI achieves a $6$~dB PAPR reduction over the original OFDM signal with $N = 256$ subcarriers, which is, to the best of our knowledge, the best reported performance among TI schemes. Scalability is also evaluated under $\mathcal{O}(N\log N)$ and $\mathcal{O}(N)$ complexity configurations, showing consistent performance gains for $N \in \{128, 256, 512\}$ in both OFDM and AFDM. Furthermore, the proposed schemes achieve superior symbol error rate (SER) performance under nonlinear PA distortion, confirming the practical significance beyond PAPR metrics.

\section{PAPR and TI}\label{sec:sys}
\subsection{PAPR of OFDM and AFDM Signals}
Consider an OFDM or AFDM system with $N$ subcarriers. Let $\mathbf{s} = [s_0, s_1, \ldots, s_{N-1}]^T \in \mathbb{C}^N$ denote a block of modulated symbols with $\ex[\mathbf{s}] = \mathbf{0}$ and $\mathbb{E}[\mathbf{s}\mathbf{s}^H]=E_s\mathbf{I}$, where $(\cdot)^T$ and $(\cdot)^H$ denote the transpose and conjugate transpose, respectively. To capture the continuous-time PAPR accurately, an oversampling factor $L$ is applied, and the $n$th time-domain sample is given by the $LN$-point inverse discrete affine Fourier transform (IDAFT):
\begin{equation}
	x_n\! = \! \frac{1}{\sqrt{N}} \sum_{k=0}^{N-1}\!s_k \, e^{j2\pi\left(\alpha_1 n^2 + \frac{kn}{LN} + \alpha_2 k^2\right)}, ~n = 0, 1, \ldots, LN-1,
	\label{eq:afdm_time}
\end{equation}
where $\alpha_1, \alpha_2 \in [0,1)$ are the chirp parameters for AFDM~\cite{Bemani2023}, and zero-padding in the frequency domain (i.e., $s_k = 0$ for $k = N, \ldots, LN-1$) is implicit. When $\alpha_1 = \alpha_2 = 0$, \eqref{eq:afdm_time} reduces to the standard OFDM inverse discrete Fourier transform (IDFT).
In matrix form, $\mathbf{x} = \mathbf{A}^H\mathbf{s}$,
where  
\begin{equation}\label{eq:A}
	\mathbf{A}^H = \mathbf{D}_t \mathbf{F}_L^H \mathbf{D}_f
\end{equation}
with $\mathbf{F}_L \in \mathbb{C}^{N \times LN}$ and $[\mathbf{F}_L]_{k,n} = e^{-j2\pi kn/(LN)}/\sqrt{N}$, $\mathbf{D}_f\in \mathbb{C}^{N\times N}$ and $\mathbf{D}_t\in\mathbb{C}^{LN\times LN}$ are diagonal matrices with  $[\mathbf{D}_f]_{k,k} = e^{j2\pi \alpha_2 k^2}$ and $[\mathbf{D}_t]_{n,n} = e^{j2\pi \alpha_1 n^2}$ ($[\cdot]_{k,n}$ denotes the $(k,n)$th entry.).

The peak-to-average power ratio (PAPR) of the time-domain signal is defined as
\begin{equation}
	\mathrm{PAPR}(\mathbf{x}) \triangleq \frac{\|\mathbf{x}\|^2_{\infty}}{\ex[ |x_n|^2]},
	\label{eq:papr_def}
\end{equation}
where $\|\mathbf{x}\|^2_{\infty} \triangleq \max_{n}|x_n|^2$.

\subsection{Tone Injection}
TI reduces PAPR by shifting selected entries in $\mathbf{s}$ in a lattice space. Each frequency-domain symbol under TI becomes $\tilde{s}_n = s_n+\delta b_n$, where $b_n\in\{R+jI\mid R,I\in\mathbb{Z}\}$ is a complex integer. $\delta$ is selected such that the receiver can recover $s_n$ from $\tilde{s}_n$ with a modulo operation given by 
\begin{equation}\label{eq:removeb}
	s_n = \tilde{s}_n - \delta\bigg\lfloor \frac{\tilde{s}_n}{\delta}+\frac{1}{2}\bigg\rfloor,
\end{equation}
where $\lfloor\cdot\rfloor$ rounds down to the nearest integer. A common selection for $M$-ary square QAM is $\delta = d\sqrt{M}$, where $d$ denotes the minimum Euclidean distance in the constellation. 

\eqref{eq:papr_def} shows that a pure increase in the denominator (i.e., the average power) also leads to PAPR reduction while having no practical benefit. Instead, minimizing the peak power 
guarantees PAPR minimization without relying on an artificial increase in 
average power, ensuring that any reduction is achieved through genuine 
suppression of the peak envelope. The problem of finding the peak-power-minimizing $\mathbf{b} = [b_0,b_1,\cdots,b_{N-1}]^T$ is formulated as
\begin{equation}\label{eq:ti_opt}
	\argmin_{\mathbf{b}} \|\mathbf{A}^H(\mathbf{s}+\delta\mathbf{b})\|^2_\infty.
\end{equation}
Solving \eqref{eq:ti_opt} to global optimality requires exhaustive search in an infinite lattice space, leading to a prohibitive computational complexity. 

For notational simplicity, $\mathbf{x}$ is reused for the time-domain signal under TI, given by
\begin{equation}\label{eq:x_ti}
	\mathbf{x} = \mathbf{A}^H(\mathbf{s}+\delta\mathbf{b}).
\end{equation}

\section{TI via Candidate Ranking}\label{sec:cr}
In this section, we first introduce the concept of local peaks of the time-domain signal and establish their utility in TI. Two novel TI schemes are then proposed, which iteratively update $\mathbf{b}$ via a closed-form candidate ranking operation based on multiple local peaks.

\subsection{Local Peaks}
\begin{definition}
	A time-domain sample $x_n$ is a \emph{local peak} of $\mathbf{x}$ if
	\begin{equation}\label{eq:lp}
		|x_n| \ge \max\!\left(|x_{(n-1)\bmod LN}|,\, |x_{(n+1)\bmod LN}|\right),
	\end{equation}
	where $\bmod$ denotes the modulo operator.
\end{definition}
The modulo-$LN$ indexing reflects the cyclic structure of OFDM/AFDM signals, so that $x_0$ and $x_{LN-1}$ are treated as adjacent samples. 

Assume $\mathbb{E}[\tilde{\mathbf{s}}] = \mathbf{0}$ and $\mathbb{E}[\tilde{\mathbf{s}}\tilde{\mathbf{s}}^H] = \sigma^2\mathbf{I}$ with $\sigma^2> E_s$. When $L = 1$ (no oversampling), the samples $\{x_n\}$ are mutually uncorrelated and, as $N \to \infty$, asymptotically independent by the CLT~\cite{Ochiai2001}. However, when $L > 1$, the oversampled transform matrix is no longer unitary and adjacent samples become correlated. As $N\to\infty$, by the CLT, $x_n \sim \mathcal{CN}(0, \sigma^2)$, so $|x_n|^2 \sim \mathrm{Exp}(\sigma^2)$. For two jointly complex Gaussian random variables $x_n$ and $x_{n+\Delta n}$ with $\Delta n \neq 0$, the covariance between their powers is
\begin{equation}\label{eq:cov}
	\begin{aligned}
		&\mathrm{Cov}\!\left(|x_n|^2, |x_{n+\Delta n}|^2\right) \\
		&= \mathbb{E}\!\left[|x_n|^2|x_{n+\Delta n}|^2\right] - \mathbb{E}\!\left[|x_n|^2\right]\mathbb{E}\!\left[|x_{n+\Delta n}|^2\right] \\
		&\overset{(\mathrm{i})}{=} \mathbb{E}\!\left[x_n x_n^*\right]\mathbb{E}\!\left[x_{n+\Delta n} x_{n+\Delta n}^*\right]
		+ \mathbb{E}\!\left[x_n x_{n+\Delta n}^*\right]\mathbb{E}\!\left[x_n^* x_{n+\Delta n}\right] \\
		&\quad + \underbrace{\mathbb{E}\!\left[x_n x_{n+\Delta n}\right]}_{=\,0}\underbrace{\mathbb{E}\!\left[x_n^* x_{n+\Delta n}^*\right]}_{=\,0} - \sigma^4\\
		&= \sigma^4 + \left|\mathbb{E}\!\left[x_n x_{n+\Delta n}^*\right]\right|^2 - \sigma^4 = \left|\mathbb{E}\!\left[x_n x_{n+\Delta n}^*\right]\right|^2\\
		&= \frac{1}{N^2}\left|\mathbb{E}\!\left[\sum_{k=0}^{N-1}\tilde{s}_k e^{j2\pi\!\left(\alpha_1 n^2+\frac{kn}{LN}+\alpha_2 k^2\right)}\right.\right.\\
		&\quad \left.\left.\cdot\sum_{l=0}^{N-1}\tilde{s}_l^* e^{-j2\pi\!\left(\alpha_1(n+\Delta n)^2+\frac{l(n+\Delta n)}{LN}+\alpha_2 l^2\right)}\right]\right|^2\\
		&= \frac{\sigma^4}{N^2}\left|\sum_{k=0}^{N-1}e^{-j2\pi k\Delta n/(LN)}\right|^2
		\overset{(\mathrm{ii})}{=} \frac{\sigma^4}{N^2}\left|\frac{1-e^{-j2\pi\Delta n/L}}{1-e^{-j2\pi\Delta n/(LN)}}\right|^2\\
		&= \sigma^4\left|\frac{\sin(\pi\Delta n/L)}{N\sin(\pi\Delta n/(LN))}\right|^2,
	\end{aligned}
\end{equation}
where $(\mathrm{i})$ follows from the Isserlis--Wick theorem for zero-mean jointly complex Gaussian pairs~\cite{Isserlis1918}, $(\mathrm{ii})$ follows from the finite geometric series formula, and $(\cdot)^*$ denotes the complex conjugate. As $N\to\infty$, \eqref{eq:cov} reduces to $\sigma^4\operatorname{sinc}^2(\Delta n/L)$, which is strictly positive for $|\Delta n|<L$. Therefore, the power of each local peak is positively correlated with its neighbouring samples within a window of $L$ samples, and can be used as proxies for the surrounding signal envelope.

\subsection{CR-TI}
Denote the index set of all local peaks as $\mathcal{P}=\{p_1,p_2,\cdots\}$, where the indices are sorted in descending order of magnitude, i.e., $|x_{p_1}|\ge|x_{p_2}|\ge\cdots$. For iterative TI, $\mathbf{b}$ is gradually updated based on the time-domain signal in each iteration. Since $|\mathcal{P}|$ is typically much smaller than $LN$, selecting TI candidates based on $\mathbf{x}_\mathcal{P}$ requires significantly lower computational complexity than operating on the full signal $\mathbf{x}$. Motivated by this observation, we propose the following CR-TI scheme.

The variables generated from the $t$th iteration of CR-TI are labeled with $(\cdot)[t]$. Initialize $\mathbf{b}[0] = \mathbf{0}$. CR-TI adopts a minimal update step in each iteration, i.e., $||\mathbf{b}[t] - \mathbf{b}[t-1]||_2 = 1$ with $\|\cdot\|_2$ denoting the Euclidean norm. Under this constraint, all candidate values for $\Delta\mathbf{b}[t]\triangleq \mathbf{b}[t] - \mathbf{b}[t-1]$ can be represented in the matrix as \begin{equation}\label{eq:barC}
	\overline{\mathbf{C}} \triangleq [\mathbf{I}~-\mathbf{I}~j\mathbf{I}~-j\mathbf{I}]\in \mathbb{C}^{N\times4N},
\end{equation}
with each column representing a unique candidate. According to \eqref{eq:x_ti}, the time-domain difference is given by $\Delta\mathbf{x}[t] \triangleq \mathbf{x}[t] - \mathbf{x}[t-1] = \delta\mathbf{A}^H\Delta\mathbf{b}[t]$. Therefore, the time-domain candidate matrix for $\Delta\mathbf{x}[t]$ is defined as 
\begin{equation}\label{eq:C}
	\mathbf{C} \triangleq \delta\mathbf{A}^H\overline{\mathbf{C}}\in\mathbb{C}^{LN\times4N},
\end{equation} 
which is only determined by system parameters.  Let $\mathcal{P}[t]$ denote the index set of the local peaks in $\mathbf{x}[t]$. The reduction in power of the $n$th local peak contributed by the $k$th candidate is given by
\begin{equation}
	\begin{aligned}
		&|x_{p_n}[t-1]|^2 - |x_{p_n}[t-1]+c_{p_n,k}|^2 \\
		=& -2\mathfrak{R}(x_{p_n}[t-1]c^*_{p_n,k})-|c_{p_n,k}|^2\\
		=& -\frac{2}{\sqrt{N}}|x_{p_n}[t-1]|\cos(\theta_{p_n}[t-1]-\phi_{p_n,k})-\frac{1}{N}\\
		\approx&-\frac{2}{\sqrt{N}}|x_{p_n}[t-1]|\cos(\theta_{p_n}[t-1]-\phi_{p_n,k}),
	\end{aligned}
\end{equation}
where $p_n$ denotes the $n$th index in $\mathcal{P}[t-1]$, $c_{p_n,k}$ denotes the entry of $\mathbf{C}$ in the $p_n$th row and $k$th column, $\theta_{p_n}[t-1] = \angle x_{p_n}[t-1]$, and $\phi_{p_n,k} = \angle c_{p_n,k}$. 
\begin{definition}
	The \emph{negated weighted cosine similarity (NWCS)} of the $n$th local peak with respect to the $k$th TI candidate in the $t$th iteration is defined as
	\begin{equation}\label{eq:wcs}
		r_{n,k}[t] \triangleq -|x_{p_n}[t-1]|^\beta\cos(\theta_{p_n}[t-1]-\phi_{p_n,k}),
	\end{equation}
	where $\beta > 0$ is an exponent controlling the emphasis on peak magnitude in the ranking.
\end{definition}
\begin{definition}
	The \emph{score} of the $k$th TI candidate in the $t$th iteration is defined as
	\begin{equation}\label{eq:score}
		R_k[t] \triangleq \sum_{n=1}^{N_p}r_{n,k}[t],
	\end{equation}
	where $0 < N_p \le |\mathcal{P}[t-1]|$ denotes the number of highest-power local peaks considered in the ranking.
\end{definition}
The $k$th candidate is valid if $R_k[t] > 0$. When valid candidates exist, CR-TI selects the $k[t]$th column of $\overline{\mathbf{C}}$ as $\Delta\mathbf{b}[t]$ according to
\begin{equation}\label{eq:ranking}
	k[t] = \argmax_{k\in\{1,2,\cdots,4N\}} R_k[t],
\end{equation}
which constitutes a ranking procedure that selects the highest-scored candidate, yielding the greatest weighted power reduction across the monitored local peaks. If no valid candidate exists, i.e., $R_k[t] \le 0$, $\forall k$, the iteration terminates and $\mathbf{b} = \mathbf{b}[t-1]$ is returned.

\begin{algorithm}[t]
	\caption{CR-TI and FCR-TI.}
	\label{alg:1}
	\begin{algorithmic}[1]
		\small
		\Statex \textbf{Input}: $N$, $L$, $\mathbf{s}$, $\delta$, $\beta$, $T$, $\alpha_1$, $\alpha_2$, $N_p$, and $N_c$ (for FCR-TI)
		\Statex \textbf{Output}: $\mathbf{b}$.
		\State Obtain $\mathbf{A}$ and $\overline{\mathbf{C}}$ according to \eqref{eq:A} and \eqref{eq:barC}.
		\State Initialize $\mathbf{b}[0] \leftarrow \mathbf{0}$ and $\mathbf{x}[0] \leftarrow \mathbf{A}^H\mathbf{s}$.
		\If{FCR-TI is used}
		\State Compute $\mathbf{f}$ via \eqref{eq:clip} and $\mathbf{g} = \mathbf{A}\mathbf{f}$.
		\State Sort $\{|g_n|\}$ in descending order; let $\mathcal{Q}=\{q_1,\cdots,q_N\}$ be the sorted indices.
		\State Set $\overline{\mathbf{C}} \leftarrow$ \eqref{eq:cf}.
		\EndIf
		\State Obtain $\mathbf{C}$ via \eqref{eq:C}.
		\For{$t=1: T$}
		\State Obtain the index set $\mathcal{P}[t-1]$ of local peaks in $\mathbf{x}[t-1]$ according to \eqref{eq:lp}.
		\State Obtain $R_k[t]$ for all $k$ according to \eqref{eq:wcs} and \eqref{eq:score}.
		\If{$R_k[t]\le 0$, $\forall k$}
		\State Update $\mathbf{b}[t]\leftarrow\mathbf{b}[t-1]$
		\State \textbf{break}
		\Else
		\State Obtain $k[t]$ via \eqref{eq:ranking}.
		\State Update $\mathbf{b}[t]\leftarrow\mathbf{b}[t-1]+\overline{\mathbf{C}}_{:,k[t]}$.
		\State Update $\mathbf{x}[t]\leftarrow \mathbf{x}[t-1]+\mathbf{C}_{:,k[t]}$.
		\EndIf
		\EndFor
		\State \textbf{Output} $\mathbf{b}\leftarrow\mathbf{b}[t]$.
	\end{algorithmic}
\end{algorithm}

\subsection{FCR-TI}
As will be demonstrated in Section~\ref{sec:comp}, the computational complexity of CR-TI is dominated by \eqref{eq:wcs}. To reduce the number of NWCSs calculated per iteration, we propose FCR-TI, which restricts the candidate set based on the clipping noise spectrum.

Following~\cite{wang2008clipping}, the $n$th time-domain clipping noise sample is defined as
\begin{equation}\label{eq:clip}
	f_n = \begin{cases}
		x_n, & |x_n|\ge \eta,\\
		0, & |x_n|<\eta,
	\end{cases}
\end{equation}
where $\eta>0$ is the clipping threshold. The in-band clipping noise spectrum is obtained via $\mathbf{g}=[g_0,g_1,\cdots,g_{N-1}]^T=\mathbf{A}\mathbf{f}$, with $\mathbf{f} = [f_0,f_1,\cdots,f_{LN-1}]^T$. A larger $|g_n|$ indicates a greater contribution of subcarrier $n$ to the clipping noise, and hence a higher potential for PAPR reduction via TI. Sorting $\{|g_n|\}$ in descending order, the corresponding subcarrier indices are collected in a set $\mathcal{Q}=\{q_1,q_2,\cdots,q_N\}$. It has been demonstrated in~\cite{Wang2016} that restricting TI to the subcarriers indexed by the leading entries of $\mathcal{Q}$ incurs only a modest PAPR penalty. Accordingly, FCR-TI restricts the candidate matrix $\overline{\mathbf{C}}$ to its submatrix formed by the columns corresponding to the first $N_c$ indices in $\mathcal{Q}$, denoted
\begin{equation}\label{eq:cf}
	\overline{\mathbf{C}}_{:,\{q_n+kN\mid n=1,\cdots, N_c,~k=0,1,2,3\}}\in\mathbb{C}^{N\times 4N_c},
\end{equation}
where $\mathbf{M}_{:,\mathcal{I}}$ denotes the submatrix of $\mathbf{M}$ formed by the columns indexed by $\mathcal{I}$. The procedure of CR-TI and FCR-TI is summarized in \textbf{Algorithm~\ref{alg:1}}.

\subsection{Integration with DFS}\label{sec:dfs}
To exploit the tree structure to explore multiple search paths in the candidate search process, we integrate DFS~\cite{CLRS} into the proposed schemes. With DFS, CR-TI and FCR-TI do not terminate immediately upon reaching a state where no valid candidate exists. Instead, the resulting output $\mathbf{b}_i$ is recorded as the $i$th \emph{leaf node} with a label $P_i = \|\mathbf{A}^H(\mathbf{s}+\delta\mathbf{b}_i)\|^2_\infty$. The search then traces back to the \emph{parent node} and continues with the next best candidate, repeating until all leaves are visited or the maximum number of iterations $T$ is reached. The final output is the $\mathbf{b}_i$ with the lowest $P_i$.

\begin{figure}[t]
	\centering
	\includegraphics[width=0.6\columnwidth]{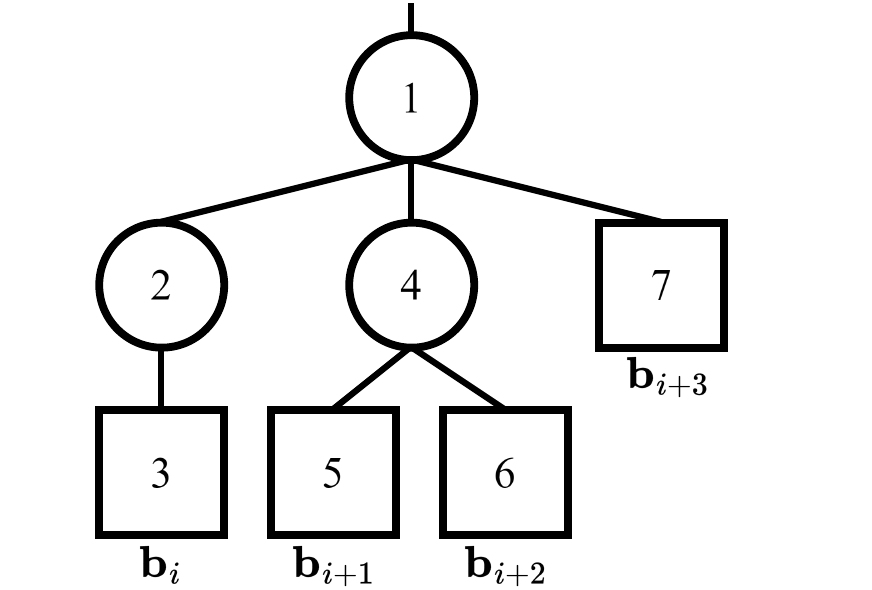}
	\caption{Illustration of the DFS procedure for CR-TI and FCR-TI. Circles represent parent nodes and squares represent leaf nodes, each corresponding to a potential output $\mathbf{b}$. Nodes are ordered left to right in descending score order within each layer, and indices denote the visitation order.}
	\label{fig:dfs}
\end{figure}

Fig.~\ref{fig:dfs} illustrates the DFS procedure. Circles represent internal nodes with valid candidates, and squares represent leaf nodes, each corresponding to a potential output $\mathbf{b}$. Within each layer, nodes are ordered from left to right in descending order of their scores. The index on each node denotes its visitation order.

\subsection{Complexity Analysis}\label{sec:comp}
The following analysis is based on \textbf{Algorithm~\ref{alg:1}}. In the initialization stage, Step~2 computes $\mathbf{x}[0] = \mathbf{A}^H\mathbf{s}$ via the FFT with complexity $\mathcal{O}(LN\log(LN))$. For FCR-TI, Step~4 additionally computes $\mathbf{g} = \mathbf{A}\mathbf{f}$ with the same complexity, and Step~5 sorts $N$ values with complexity $\mathcal{O}(N\log N)$, which is dominated by the former. Since $\mathbf{A}$ and the candidate matrices $\mathbf{C}$ and $\overline{\mathbf{C}}$ are constant for a given system configuration, Steps~6 and~8 reduce to simple matrix slicing with negligible cost. The initialization stage therefore incurs an overall complexity of $\mathcal{O}(LN\log(LN))$.

Within the main loop, Step~10 performs a full scan of the $LN$ time-domain samples to identify local peaks, followed by a partial sort to retain the top $N_p$ entries, yielding a complexity of $\mathcal{O}(LN\log N_p)$. Step~11 computes $4N_p N_c$ NWCSs, contributing $\mathcal{O}(N_p N_c)$. The remaining steps involve scalar updates or index lookups with negligible overhead. The worst-case per-iteration complexity is therefore $\mathcal{O}(LN\log N_p + N_p N_c)$, and the overall complexity of the proposed schemes is
\begin{equation}
	\mathcal{O}\!\left(LN\log(LN) + T\left(LN\log N_p + N_p N_c\right)\right). \label{eq:complexity}
\end{equation}
The upper bound is obtained by setting $N_p = |\mathcal{P}|$ and $N_c = N$. Since $|\mathcal{P}|$ is typically $\mathcal{O}(N)$, the complexity of CR-TI is upper-bounded by $\mathcal{O}(TN^2)$. For FCR-TI, selecting $N_p$ and $N_c$ such that $N_p N_c = \mathcal{O}(LN\log N_p)$ balances the two loop terms, reducing the overall complexity to
\begin{equation}
	\mathcal{O}\!\left(LN\log(LN) + TLN\log N_p\right), \label{eq:complexity_cf}
\end{equation}
which scales near-linearly in $N$ for fixed $L$, $T$, and $N_p$, comparable to FFT.

Table~\ref{tab:1} compares the overall complexity of the proposed schemes with cross-entropy-based TI (CE-TI)~\cite{Hu2021} and DS-TI~\cite{Wang2016}. In the literature, $T_{\mathrm{CE}}U$ and $T_{\mathrm{DS}}V$ are typically $\mathcal{O}(N)$. As shown in Section~\ref{sec:sim}, $T$ between $10$ and $40$ suffices for the proposed schemes, whose complexity upper bound is therefore comparable to CE-TI and DS-TI, and can be significantly reduced by tuning $N_p$ and $N_c$.

\begin{table}[t]
	\centering
	\caption{Complexity Comparison of TI Schemes}
	\label{tab:1}
	\renewcommand{\arraystretch}{1.3}
	\begin{tabular}{lc}
		\hline\hline
		\textbf{TI Scheme} & \textbf{Complexity ($\mathcal{O}(\cdot)$)}\\
		\hline
		CE-TI~\cite{Hu2021}          & $T_{\mathrm{CE}}ULN\log(LN)$                      \\
		DS-TI~\cite{Wang2016}          & $T_{\mathrm{DS}}VLN$                              \\
		CR/FCR-TI (general)      & $LN\log(LN) + T(LN\log N_p + N_p N_c)$           \\
		CR/FCR-TI (upper bound)  & $TN^2$                                            \\
		FCR-TI (near-linear)     & $LN\log(LN) + TLN\log N_p$                       \\
		\hline\hline
	\end{tabular}
\end{table}

\section{Simulation Results}\label{sec:sim}
Simulations are conducted for OFDM and AFDM with $c_1 = 1/(2N)$ and $c_2 = 0$ under the signal model in Section~\ref{sec:sys}. An oversampling factor $L = 8$ is used for accurate peak capturing.  64-QAM is adopted throughout as it is more sensitive to nonlinear distortion than lower-order modulations. $\beta = 4$ and DFS are applied to both proposed schemes, with $T$ being the maximum number of iterations across the entire DFS. 

For OFDM, the proposed schemes are compared against:
\begin{enumerate}
	\item OFDM without TI, as the unoptimized benchmark;
	\item CE-TI~\cite{Chen2010}, a classic probability-based iterative TI technique. The implementation in~\cite{Hu2021} is adopted. The number of iterations and randomly generated candidates are set to $T_{\mathrm{CE}} = 10$ and $U = 32$, respectively.
	\item DS-TI~\cite{Wang2016}, which represents the state-of-the-art in complexity-performance trade-off among TI schemes, with $V = 3$ equivalent constellation points and $T_{\mathrm{DS}} \in \{16, 64\}$.
\end{enumerate}
The clipping noise threshold $\eta$ is set to $\sqrt{10^{4/10}E_s}$ for DS-TI and $\sqrt{10^{5/10}E_s}$ for FCR-TI, corresponding to best empirical performance.
\subsection{PAPR Performance vs. Baseline Schemes in OFDM}
\begin{figure}[t]
	\centering
	\includegraphics[width=0.9\columnwidth]{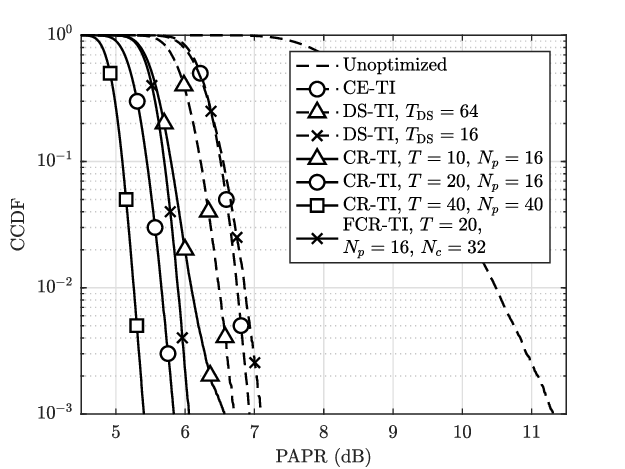}
	\caption{CCDF of PAPR for OFDM with $N = 256$, comparing CR-TI and FCR-TI against CE-TI, DS-TI, and unoptimized OFDM.}
	\label{fig:papr256}
\end{figure}

Fig.~\ref{fig:papr256} shows the PAPR CCDF for OFDM with $N = 256$. All proposed schemes outperform the baselines. CR-TI with $T = 20$ and $N_p = 16$ achieves approximately $1$~dB gain over DS-TI ($T_{\mathrm{DS}} = 64$) at comparable complexity (Table~\ref{tab:1}). Increasing to $T = 40$ and $N_p = 40$ further reduces the $10^{-3}$ PAPR to $5.4$~dB, corresponding to a $6$~dB reduction over unoptimized OFDM. To the best of our knowledge, this is the best reported PAPR performance for $N = 256$ among TI techniques. Reducing to $T = 10$ causes notable degradation at the tail of the CCDF, suggesting $T = 20$ as the preferred trade-off. FCR-TI with $T = 20$, $N_p = 16$, and $N_c = 32$ achieves a $10^{-3}$ PAPR of $6$~dB, outperforming DS-TI ($T_{\mathrm{DS}} = 64$) by $0.6$~dB at a significantly lower complexity, comparable to that of DS-TI ($T_{\mathrm{DS}} = 16$).

\begin{figure}[t]
	\centering
	\subfloat[OFDM]{
		\includegraphics[width=0.9\columnwidth]{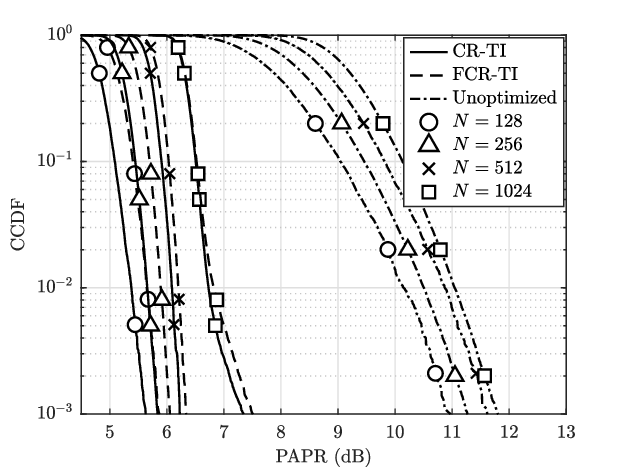}
		\label{fig:papr_ofdm}
	}\\
	\subfloat[AFDM]{
		\includegraphics[width=0.9\columnwidth]{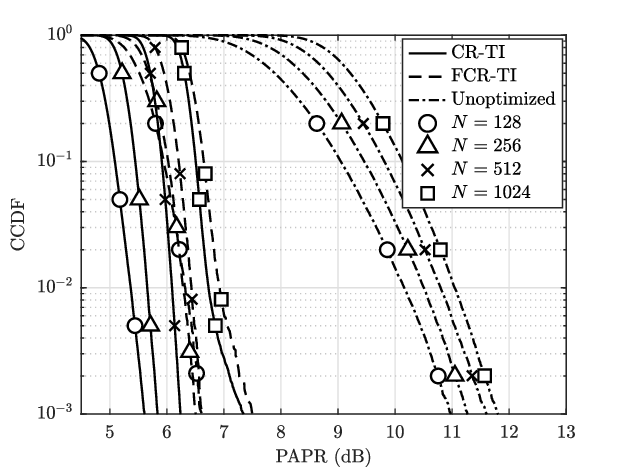}
		\label{fig:papr_afdm}
	}
	\caption{CCDF of PAPR for CR-TI and FCR-TI against unoptimized signals with $N \in \{128, 256, 512, 1024\}$ in (a) OFDM and (b) AFDM.}
	\label{fig:papr_N}
\end{figure}

\subsection{PAPR Performance for Varying $N$}
We evaluate the scalability of the proposed schemes over $N \in \{128, 256, 512, 1024\}$, with $T = 20$, $N_p = 2\log N$, and $N_c = \lfloor NL/(8\log N)\rceil$ ($\lfloor\cdot\rceil$ denotes rounding to the nearest integer) for FCR-TI, so that the per-iteration NWCS count satisfies $4N_c N_p \approx NL$. Under this configuration, the per-iteration complexity of CR-TI and FCR-TI scales as $\mathcal{O}(N\log N)$ and $\mathcal{O}(N)$, respectively (Table~\ref{tab:1}).

Fig.~\ref{fig:papr_ofdm} shows the OFDM results. For $N \le 512$, both schemes exhibit less than $0.5$~dB PAPR increase as $N$ doubles, closely tracking the scaling of the unoptimized OFDM signal and demonstrating good scalability. At $N = 1024$, the degradation exceeds this trend, suggesting that the parameters require further tuning for wideband configurations.

Fig.~\ref{fig:papr_afdm} shows the AFDM results. While CR-TI maintains similar scalability to the OFDM case, FCR-TI exhibits a notably larger PAPR increase relative to CR-TI at smaller $N$. In particular, at $N = 128$, the gap between FCR-TI and CR-TI widens from approximately $0.3$~dB to $1$~dB, yielding nearly identical FCR-TI performance to $N = 256$. This suggests that the clipping noise behaves differently in AFDM than in OFDM, and that FCR-TI requires further development for AFDM systems.
\subsection{SER Performance}
We evaluate the SER performance for $N=256$ under PA distortion modelled by a soft limiter. The $n$th time-domain sample after the soft limiter is
\begin{equation}
	\overline{x}_n = \begin{cases}
		x_n, & |x_n| < \eta',\\
		\eta' e^{j\angle x_n}, & |x_n| \ge \eta',
	\end{cases}
\end{equation}
where $\eta' = \sqrt{10^{4.5/10}E_s}$ is the clipping amplitude threshold set to $4.5$~dB above the average signal power. An AWGN channel is assumed, and the frequency-domain received signal is $\mathbf{y} = \mathbf{A}\overline{\mathbf{x}} + \mathbf{w}$, where $\mathbf{w} \sim \mathcal{CN}(\mathbf{0}, N_0\mathbf{I})$. The receiver performs per-subcarrier detection on $\mathbf{y}$ after removing the injected integers via~\eqref{eq:removeb}. For fair comparison, $E_s$ is normalized by the average transmit power increase introduced by TI, listed in Table~\ref{tab:2}.

\begin{figure}[t]
	\centering
	\subfloat[OFDM]{
		\includegraphics[width=0.9\columnwidth]{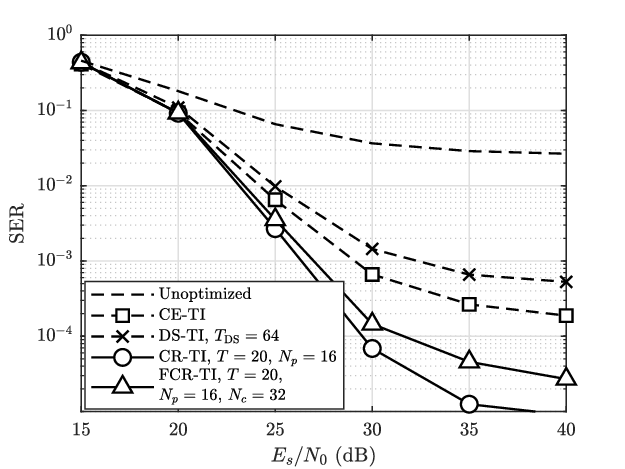}
		\label{fig:ser_ofdm}
	}\\
	\subfloat[AFDM]{
		\includegraphics[width=0.9\columnwidth]{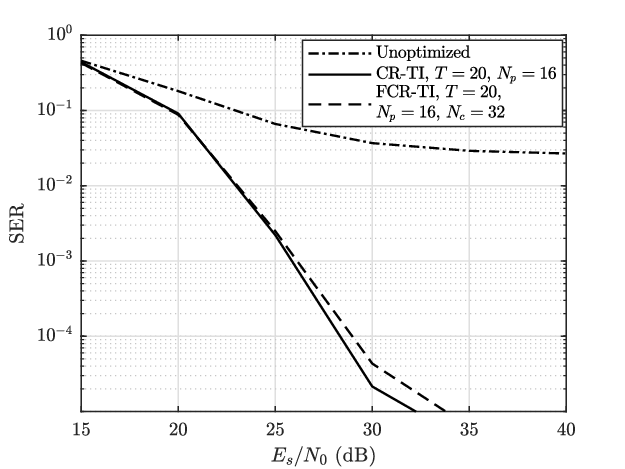}
		\label{fig:ser_afdm}
	}
	\caption{SER versus $E_s/N_0$ for CR-TI and FCR-TI against CE-TI, DS-TI, and unoptimized signals under a soft limiter with $4.5$~dB clipping threshold in (a) OFDM and (b) AFDM.}
	\label{fig:ser}
\end{figure}

Fig.~\ref{fig:ser_ofdm} shows the SER versus $E_s/N_0$ for OFDM. Without TI, the original OFDM signal exhibits a SER floor at $3\times10^{-2}$ due to high PAPR. CR-TI and FCR-TI suppress this floor to below $3\times10^{-5}$ and $1\times10^{-5}$, respectively, representing over three orders of magnitude improvement over the unoptimized signal and approximately one order of magnitude below the state-of-the-art TI baselines. At $10^{-3}$ SER, the proposed schemes achieve more than $2$~dB and $5$~dB signal-to-noise-ratio gains over CE-TI and DS-TI, respectively. Interestingly, CE-TI outperforms DS-TI in SER despite exhibiting higher PAPR, likely due to its lower transmit power increase and stricter constellation extension constraints. Fig.~\ref{fig:ser_afdm} shows the SER for AFDM, where the proposed schemes maintain similar performance gains.
\begin{table}[t]
	\centering
	\caption{Average Transmit Power Increase Introduced by TI}
	\label{tab:2}
	\renewcommand{\arraystretch}{1.3}
	\begin{tabular}{lc}
		\hline\hline
		\textbf{TI Scheme} & \textbf{Power Increase (dB)}\\
		\hline
		CE-TI ($T_{\mathrm{CE}} = 10$, $U = 32$)          & $0.1$ \\
		DS-TI ($T_{\mathrm{DS}} = 64$, $V = 3$)           & $0.4$ \\
		CR-TI ($T = 20$, $N_p = 16$)$^\dagger$            & $0.6$ \\
		FCR-TI ($T = 20$, $N_p = 16$, $N_c = 32$)$^\dagger$ & $0.4$ \\
		\hline\hline
	\end{tabular}
	\footnotesize{$^\dagger$Values are identical for OFDM and AFDM.}
\end{table}
\section{Conclusion}
In this paper, two novel low-complexity TI schemes, CR-TI and FCR-TI, are proposed for OFDM and AFDM systems. Developed based on the correlation between local peaks and the full waveform, the candidate ranking procedure accurately selects the most effective TI candidate in each iteration while maintaining a low complexity. FCR-TI further reduces the complexity to FFT-level by pre-filtering the candidates based on the clipping noise. By exploiting the tree structure of the search process, DFS is further introduced to enhance PAPR performance. Simulation results demonstrate that the proposed schemes achieve over $1$~dB PAPR gain over state-of-the-art TI baselines at comparable complexity, with consistent performance across varying numbers of subcarriers in both OFDM and AFDM. Superior SER performance under nonlinear PA distortion further confirms the practical significance of the proposed schemes. Future work will investigate the degraded performance of FCR-TI in AFDM systems and optimal parameter configurations for large-scale systems with $N \ge 1024$.
\balance
\bibliographystyle{IEEEtran}
\bibliography{pimrc2026}

\end{document}